\documentclass[12pt]{article}
\begin{document}

\begin{center}
{\large \bf Deuteron disintegration by reactor antineutrinos}
\vspace{0.5 cm}

\begin{small}
\renewcommand{\thefootnote}{*}
L.M.Slad\footnote{slad@theory.sinp.msu.ru} \\
{\it Skobeltsyn Institute of Nuclear Physics,
Lomonosov Moscow State University, Moscow 119991, Russia}
\end{small}
\end{center}

\vspace{0.3 cm}

\begin{footnotesize}
The existence of a new interaction involving the electron neutrino and the nucleons, which has received a convincing confirmation through a good agreement between the theoretical and experimental results concerning all observable processes with solar neutrinos, should also inevitably manifest itself in the deuteron disintegration by reactor antineutrinos neutral currents. In this paper, the analytical and numerical characteristics of such a disintegration are presented. The attention is drawn to the problem of finding the neutron registration efficiency, discussed in the preparation of the experiment at the Sudbury Neutrino Observatory and in a number of special studies, to the role of quenching gas in proportional counters with helium-3 and to three performed reactor experiments with heavy water.
\end{footnotesize}

\vspace{0.5 cm}

\begin{small}

\begin{center}
{\large \bf 1. Introduction}
\end{center}

In the work \cite{1}, we have presented a substantiation of the existence of a hidden enough new, semi-weak, interaction involving at least the electron neutrino and the nucleons, which is carried by an (almost) massless pseudoscalar boson $\varphi_{ps}$. The electron does not possess such an interaction at the tree level. We adhere to the classical concept that the neutrino field of any sort is described by a bispinor representation of the proper Lorenz group and obeys the Dirac equation. The semi-weak interaction Lagrangian has the following form
\begin{equation}
{\cal L} = ig_{\nu_{e}ps}\bar{\nu}_{e}\gamma^{5}\nu_{e}\varphi_{ps}+
ig_{Nps}\bar{p}\gamma^{5}p\varphi_{ps}-ig_{Nps}\bar{n}\gamma^{5}n\varphi_{ps}.
\label{1}
\end{equation}

The semi-weak interaction manifests itself clearly in the results of experiments with solar neutrinos due to the fact that the Sun essentially plays the role of part of the experimental setup, providing about 10 collisions with its nucleons for every neutrino. Each such a collision, firstly, changes the neutrino handedness, so that at the exit from the Sun the fluxes of left- and right-handed electron neutrinos are approximately equal, and, secondly, reduces the neutrino energy $\omega$ in proportion to its square: $\Delta \omega \simeq \omega^{2}/M$, where $M$ is the nucleon mass. Having the only free parameter, the effective number of neutrino collisions with the Sun nucleons $n_{0}$, or the value of the product of the coupling constants $g_{ps\nu_{e}}g_{psN}$, we obtained a good agreement between the theoretical results based on Lagrangian (\ref{1}) and the results of all the five types of experiments with solar neutrinos: ${}^{37}{\rm Cl} \rightarrow {}^{37}{\rm Ar}$, ${}^{71}{\rm Ga} \rightarrow {}^{71}{\rm Ge}$, $\nu_{e} e^{-}\rightarrow \nu_{e} e^{-}$, $\nu_{e}D \rightarrow  e^{-}pp$, and $\nu_{e}D \rightarrow \nu_{e} np$. Such an agreement is reached at $n_{0}=11$. On the basis of this number, the following  estimate is obtained
\begin{equation}
\frac{g_{\nu_{e}ps}g_{Nps}}{4\pi} = (3.2 \pm 0.2)\times 10^{-5}.
\label{2}
\end{equation}

The results of the first four mentioned experiments reflect the semi-weak interaction of the electron neutrinos in the Sun. At the same time, the results of the experiment on the deuteron disintegration by the solar neutrinos neutral currents reflect the semi-weak interaction of electron neutrinos, both in the Sun and in the terrestrial installation. Specifically, the disintegration of the deuteron into a proton and a neutron is caused by two non-interfering sub-processes. In the first sub-process, which has a standard description based on the Weinberg-Salam model, only the left-handed solar neutrinos are involved that interact with the deuteron nucleons through the $Z$-boson exchange. Accurate calculations of the cross-section of this sub-process are given in works \cite{2}--\cite{4}. In the second sub-process caused by the semi-weak interaction, both the left- and right-handed solar  neutrinos are involved due to the massless pseudoscalar boson exchange with the deuteron nucleons. The cross-section of this sub-process $\sigma_{\rm ps} (\omega)$ is calculated in work \cite{1} within a long-standing practical approximation. It is given by the formula
$$\sigma_{\rm ps}(\omega) = \frac{(g_{\nu_{e}ps}g_{Nps})^{2}}
{16\pi^{2}M^{2}}\cdot \left( \frac{M_{n}-M_{p}}{M}\right)^{2}\cdot 
\frac{\sqrt{B}(\sqrt{B}-(a_{s}\sqrt{M})^{-1})^{2}}{\omega}$$
\begin{equation}
\times \int_{0}^{\omega - B} dE_{r}\frac{(\omega - B -E_{r})
\sqrt{E_{r}}}{(E_{r}+B)^{2}(E_{r}+(a_{s}^{2}M)^{-1})},
\label{3}
\end{equation}
where $B = 2.2246$ MeV is the binding energy of the deuteron, $\delta = M_{p}-M_{n} = -1.2933$ MeV is the mass difference between the proton and the neutron, and $(a_{s}^{2}M)^{-1} = 0.0738$ MeV. The errors related to the quantity $\sigma_{\rm ps} (\omega)$ (\ref{3}) are of two types, namely, the ones coming from an error in the coupling constant product (\ref{2}) and the other from the approximation in the calculation of this quantity. We estimate the latter error from the discrepancy between the cross-section values for the $Z$-boson exchange sub-process obtained by accurate computations in work \cite{2}, and the values obtained by a method analogous to our present approximation. This is reflected in table 5 of the work \cite{1}, where the most significant difference is at the process threshold.

\begin{center}
{\large \bf 2. The rates of the deuteron disintegration by neutral and charged currents of reactor antineutrinos, $\bar{\nu}_{e}D \rightarrow \bar{\nu}_{e} np$ and $\bar{\nu}_{e}D \rightarrow  e^{+}nn$}
\end{center}
 
Note, that at the energy values $\omega$, typical for solar neutrinos and reactor antineutrinos, the cross-sections of the deuteron disintegration by neutrino and by antineutrino caused by the $Z$-boson exchange, $\sigma_{\rm Z}^{\nu_{e}}(\omega)$ or $\sigma_{\rm Z}^{\bar{\nu}_{e}}(\omega)$, show only a minor difference \cite{3}. The deuteron disintegration caused by the $\varphi_{ps}$-boson exchange has identical cross-sections for neutrinos and antineutrinos. The numerical values of the cross-sections for the both sub-processes with reactor antineutrinos are given in table 1.
\begin{center}
{\bf Table 1.} The cross-sections of the deuteron \\ disintegration
$\sigma_{\rm Z}^{\bar{\nu}_{e}}(\omega)$ É $\sigma_{\rm ps}(\omega)$ (in units of $10^{-43}$ cm$^{2}$).
\begin{tabular}{ccccccccc}
\\ 
\hline
\multicolumn{1}{c}{$\omega$}
&\multicolumn{1}{c}{$\sigma_{\rm Z}^{\bar{\nu}_{e}}(\omega)$}  
&\multicolumn{1}{c}{$\sigma_{\rm ps}(\omega)$}
&\multicolumn{1}{c}{$\omega$}
&\multicolumn{1}{c}{$\sigma_{\rm Z}^{\bar{\nu}_{e}}(\omega)$}  
&\multicolumn{1}{c}{$\sigma_{\rm ps}(\omega)$}
&\multicolumn{1}{c}{$\omega$}
&\multicolumn{1}{c}{$\sigma_{\rm Z}^{\bar{\nu}_{e}}(\omega)$}  
&\multicolumn{1}{c}{$\sigma_{\rm ps}(\omega)$} \\
(MeV) & \cite{4} & Eq. (\ref{3}) &(MeV) & \cite{4} & Eq. (\ref{3})
&(MeV) & \cite{4} & Eq. (\ref{3}) \\
\hline
2.2  & 0.0000 & 0.000 & 4.2 & 0.396 & 1.476 & 6.2 & 2.215 & 2.646 \\
2.4  & 0.0004 & 0.037 & 4.4 & 0.505 & 1.621 & 6.4 & 2.490 & 2.734 \\
2.6  & 0.0042 & 0.151 & 4.6 & 0.630 & 1.759 & 6.6 & 2.782 & 2.818 \\
2.8  & 0.0144 & 0.303 & 4.8 & 0.707 & 1.890 & 6.8 & 3.092 & 2.899 \\
3.0  & 0.0332 & 0.473 & 5.0 & 0.927 & 2.015 & 7.0 & 3.419 & 2.976 \\
3.2  & 0.0621 & 0.649 & 5.2 & 1.100 & 2.134 & 7.2 & 3.764 & 3.050 \\
3.4  & 0.1025 & 0.825 & 5.4 & 1.289 & 2.247 & 7.4 & 4.126 & 3.120 \\
3.6  & 0.1553 & 0.997 & 5.6 & 1.495 & 2.354 & 7.6 & 4.506 & 3.188 \\
3.8  & 0.2213 & 1.163 & 5.8 & 1.718 & 2.456 & 7.8 & 4.904 & 3.253 \\
4.0  & 0.3012 & 1.323 & 6.0 & 1.958 & 2.553 & 8.0 & 5.320 & 3.315 \\
\hline
\end{tabular}
\end{center}

In what follows, any quantity $A$ related to reactor antineutrinos produced in the fission of the isotopes $^{235}{\rm U}$, $^{238}{\rm U}$, $^{239}{\rm Pu}$, and $^{241}{\rm Pu}$ will be denoted, respectively, as $A_{5}$, $A_{8}$, $A_{9}$, and $A_{1}$. Recall a number of known things.  The energy released in the fission of any of the listed isotopes, $E$ (in MeV/fission), is approximately the same: $E_{5}=201.9$, $E_{8}=205.5$, $E_{9}=210.0$, and $E_{1}=213.6$ (see, for example, \cite{5}). The relative contributions  $\alpha(t)$ of the isotopes to the number of fissions in a reactor change with time. For example, during a standard operation period of a VVER-1000 reactor, they change in the intervals as follows \cite{6}: $\alpha_{5}(t) \in [0.65, 0.48]$, $\alpha_{8}(t) \in [0.07, 0.07]$, $\alpha_{9}(t) \in [0.24, 0.37]$, and $\alpha_{1}(t) \in [0.04, 0.08]$.

On the basis of table 1, it is easy to conclude that the contribution of semi-weak interaction to the rate of the deuteron disintegration by reactor antineutrinos is comparable with that of electroweak interaction. Consequently, in the theoretical calculations for such a rate, it is acceptable at the present stage to ignore inaccuracy (essential for refined experiments) in the existing description of the spectra of reactor antineutrinos $S_{i}(\omega)$, $i = 5, 8, 9, 1$, discovered by the collaboration Daya Bay \cite{7} under comparing their experimental results with the results of the widely known works \cite{8} and \cite{9}.  In our calculations, we use the results of work \cite{8}.

The cross-sections of the deuteron disintegration by neutral currents integrated over the spectra of reactor antineutrinos per one fission of a considered isotope (named as the integral cross-sections),
\begin{equation}
\Sigma_{{\rm Z},i} = \int_{B}^{8 {\rm MeV}} S_{i}(\omega) \sigma_{\rm Z}^{\bar{\nu}_{e}}(\omega) d\omega, \quad \Sigma_{{\rm ps},i} = \int_{B}^{8 {\rm MeV}} S_{i}(\omega) \sigma_{\rm ps}(\omega) d\omega,
\label{4}
\end{equation}
are presented in table 2 for both sub-processes, as well as their sums $\Sigma_{{\rm Z+ps},i}=\Sigma_{{\rm Z},i}+\Sigma_{{\rm ps},i}$. Values, presented for comparison in the last line of table 2, are the integral cross-sections of the deuteron disintegration by the reactor antineutrino charged current, calculated with using the $\sigma_{\rm cc}^{\bar{\nu}_{e}}(\omega)$ values taken from \cite{2}.
\begin{center}
{\bf Table 2.} The integral cross-sections \\ of the deuteron disintegration (in unit of \\ $10^{-43}$ cm$^{2}$/fission). \\
\begin{tabular}{ccccc}
\\ 
\hline
\multicolumn{1}{c}{}
&\multicolumn{1}{c}{$^{235}{\rm U}$}
&\multicolumn{1}{c}{$^{238}{\rm U}$}  
&\multicolumn{1}{c}{$^{239}{\rm Pu}$}
&\multicolumn{1}{c}{$^{241}{\rm Pu}$} \\
\hline
$\Sigma_{{\rm Z},i}$ & 0.328 & 0.544 & 0.191 & 0.282  \\
$\Sigma_{{\rm ps},i}$ & 1.026 & 1.586 & 0.675 & 0.948  \\
$\Sigma_{{\rm Z+ps},i}$ & 1.354 & 2.131 & 0.866 & 1.230  \\
$\Sigma_{{\rm cc},i}$ & 0.112 & 0.197 & 0.056 & 0.086  \\
\hline
\end{tabular}
\end{center}

The integral cross sections averaged over the reactor isotopes relative contributions
\begin{equation}
\Sigma^{r}_{\rm X}(t) = \alpha_{5}(t)\Sigma_{{\rm X},5}+\alpha_{8}(t)\Sigma_{{\rm X},8}+\alpha_{9}(t)\Sigma_{{\rm X},9}+\alpha_{1}(t)\Sigma_{{\rm X},1},
\label{5}
\end{equation}
where ${\rm X = Z, ps, Z+ps, cc}$, will be called the weighted integral cross-sections. During a standard operation period of a VVER-1000 reactor, the weighted integral cross-sections change within the following intervals (in unit of
$10^{-43}$ cm$^{2}$/fission): $\Sigma^{r}_{\rm Z}(t) \in [0.31, 0.29]$, $\Sigma^{r}_{\rm ps}(t) \in [0.98, 0.93]$, $\Sigma^{r}_{\rm Z+ps}(t) \in [1.29, 1.22]$, and $\Sigma^{r}_{\rm cc}(t) \in [0.103, 0.095]$. Evidently, the listed variations are too small to have an effect on the conclusion about a significant contribution of semi-weak interaction to the deuteron disintegration rate by reactor antineutrinos. In the following, we will be using the central values of these intervals with the error bands steming, respectively, from the uncertainty in the knowledge of the isotopes relative contributions, from the uncertainty in the coupling constants (\ref{2}) and from the approximation in calculating the cross-section with the boson $\varphi_{ps}$ exchange (in units of $10^{-43}$ cm$^{2}$/fission):  
\begin{eqnarray}
& & \Sigma^{r}_{\rm Z} = 0.299 \cdot (1\pm 0.032), \nonumber \\
& & \Sigma^{r}_{\rm ps} = 0.954 \cdot (1\pm 0.025) \cdot (1\pm 0.125) \cdot (1 \pm 0.057)= 0.954 \cdot (1 \pm 0.21), \nonumber \\
& & \Sigma^{r}_{\rm Z+ps} = 1.252 \cdot (1\pm 0.027) \cdot (1\pm 0.095) \cdot (1 \pm 0.043)= 1.25 \cdot (1 \pm 0.15), \nonumber \\
& & \Sigma^{r}_{\rm cc} = 0.099 \cdot (1\pm 0.040). \label{6}
\end{eqnarray}
A comparison of these cross-sections with those in table 2 for the pure uranium-235 reactor shows that the ratio between them is from 1.08 to 1.13. Since such a difference is not essential for our problem, then, in the future, we use the values of the cross-sections (\ref{6}) without any reservations. 

Take $E=205$ MeV/fission as the energy released in the reactor at the fission of any isotope. Then, for an experimental installation filled with heavy water ${\rm D}_{2}{\rm O}$ of mass $m$ and placed at the distance $R$ from the reactor with thermal power $W$, the per day number of the deuteron disintegration events $N_{\rm X}$ caused by this or that process or sub-process $X$ can be calculated by the following formula
\begin{equation}
N_{\rm X} = 1.58\cdot 10^{47} \cdot \Sigma^{r}_{\rm X} \cdot \frac{W}{\rm megawatt}\cdot \frac{m}{\rm kg} \cdot \frac{1}{4\pi R^{2}} \cdot \frac{\rm fission}{\rm day}.
\label{7}
\end{equation}

\begin{center}
{\large \bf 3.On the neutron registration problem}
\end{center}

The new experiments on the deuteron disintegration by reactor antineutrinos, which I call for in this paper, can play a very important role in confirming (or disproving) the hypothesis of the existence of a new interaction involving the electron neutrinos, if their results will have approximately the same level of reliability as the experiment on the deuteron disintegration by solar neutrinos, carried out at the Sudbury Neutrino Observatory (SNO).

At the SNO, three different techniques were used to register neutrons in events caused by neutral neutrino currents, $\nu_{e}+D \rightarrow \nu_{e}+n+p$. In the first phase of this experiment \cite{10}, the neutrons were detected via capturing by deuterium, with the emission of gamma quantum with the energy of 6.26 MeV. In the second phase \cite{11}, the NaCl salt was dissolved in heavy water and the neutron was detected by capturing it with the $^{35}{\rm Cl}$ nucleus emitting a cascade of gamma quanta with total energy of 8.6 MeV. In the third phase, the neutron was registered by proportional counters with $^{3} {\rm He}$ gas \cite{12}. The registration of neutrons in the third phase is much more complex case than in the first and second phases. Considerable preliminary studies were devoted to the development of its methodology, some of which are presented in a voluminous article \cite{13}. The absence of any noticeable defects in this technique is evidenced by the agreement between the results of all three phases of the SNO experiment.

The deuteron disintegration by reactor antineutrinos have been studied in three performed experiments. One such experiment took place in two stages at the Savannah River Plant reactor \cite{14}, \cite{15}, and another experiment with the same (or identical) installation, with the same methodology and with a number of common participants took place at the Bugey reactor-5 \cite{16}. The third experiment in two stages was carried out at the Rovno reactor \cite{17}, \cite{18}. In all these experiments, proportional counters with the $^{3}{\rm He}$ gas were used to register neutrons. The identification of events is complicated compared with the SNO experiments by the fact that, at the deuteron disintegration by reactor antineutrinos, neutrons are produced under the effect of both the neutral and charged currents, $\bar{\nu}_{e}+D \rightarrow \bar{\nu}_{e}+n+p$ and $\bar{\nu}_{e}+D \rightarrow e^{+}+n+n$. 

Published reports about each of these experiments contain some mystery, unexplained elements. We shall note some of them only so that the preparation for setting new experiments on the deuteron disintegration by reactor antineutrinos, if they will take place, was conducted at a much higher level than in the performed experiments. Below we will pay attention, firstly, to different aspects of the problem of finding the efficiency of neutron registration by helium-3 counters, which have been discussed at the preparation of the SNO experiment and in a number of special studies carried out after 2000, and, secondly, to the problem of preventing spurious pulses in the proportional counters, solved at the SNO.

At the SNO, the efficiency of neutron registration by $^{3}{\rm He}$-counters of the installation was determined by various methods described in detail in the article \cite{19}. One method is based on the uniformly distributed introduction into a ball with heavy water of the $^{24}$Na isotope with a half-life of 14.96 hours, emitting gamma rays, some of which has caused the disintegration of deuterons and the appearance of neutrons. It has gave to the neutron registration efficiency the value equal to $0.211 \pm 0.007$. Another method was in placing the $^{252}$Cf- and AmBe-neutron sources at different points of the D$_{2}$O volume using a manipulator. This method did not play an independent role, but was used to tune Monte Carlo calculations, which have established the efficiency value equal to $0.210 \pm 0.003$.

Note now that the proportional counters at the SNO are filled with a mixture of $^{3}{\rm He}$ and CF$_{4}$ in the ratio of 85$\%$ and 15$\%$ (by pressure) at 2.5 atm \cite{19}. CF$_{4}$ acts as a quenching gas, which preventes spurious pulses in the counters [13].In the work \cite{16}, the counters are filled with a mixture of $^{3}{\rm He}$ and Ar with partial pressures of 1 and 1.7 atm, respectively. In the work \cite{18}, the $^{3}{\rm He}$-gas pressure is 4 atm. In experiments \cite{15}, \cite{16}, and \cite{18}, there seems to be no quenching gas in the proportional counters.

In the meantime, I note a few facts concerning the neutron capture process $n+^{3}{\rm He} \rightarrow p+^{3}{\rm H}$. The dependence of the cross-section of this process $\sigma$ on the kinetic energy $E$ neutron is given by the following relation \cite{20}:
\begin{equation}
\sigma = (847.5 \pm 1.5) (E/{\rm eV})^{-1/2} b, \qquad {\rm if} \ E<11 \ {\rm eV}.
\label{8}
\end{equation}
From here, at the thermal energy of $E= 0.0253$ eV corresponding to the temperature $21^{\rm o}$C, one obtains that $\sigma = 5300$ b. When partial pressure of $^{3}{\rm He}$ gas is equal to 1 atm, the path length of the neutron until its capture is approximately 7.0 cm, and, at a pressure of 2.1 atm (as at the SNO), it is approximately equal to 3.3 cm. Meanwhile, the diameters of the counters with $^{3}{\rm He}$ in works \cite{15}, \cite{16}, and \cite{19} are the same and equal to 5.08 cm. The neutron, once in a proportional counter, can make there Brownian motion upto the capture by the helium-3, but can also with high probability due to its Brownian motion go beyond the counter back into heavy water. The compiler of the Monte Carlo program should take into account, if he can, this circumstance in calculating the efficiency of neutron registration. Without taking into account the possibility of neutron output from a counter, the Monte Carlo program will give an increased value of efficiency.

In the works \cite{21} and \cite{22}, an experimental study of the efficiency of neutron registration by a separate $^{3}{\rm He}$-counter at well-defined locations of AmLi- and $^{252}$Cf-neutron sources, respectively, was carried out. In the work \cite{21}, the dependence of the relative efficiency on the partial pressure of the $^{3}{\rm He}$ gas in the range from 2 to 10 atm is found. In particular, when the pressure decreases from 4 to 2 atm, the efficiency decreases by about 1.4 times. In the work \cite{22} the dependence of the absolute efficiency on the partial pressure in the range from 1.0 to 3.0 atm, at which it increases by 1.55 times, from 0.076 to 0.118, is given.

In the studies \cite{21} and \cite{22}, CO$_{2}$ is used as a quenching gas in the $^{3}{\rm He}$-counters.

\begin{center}
{\large \bf 4. A view of the performed experiments on the deuteron disintegration by reactor antineutrinos}
\end{center}

Let us consider now the lessons taken from three experiments on the deuteron disintegration by reactor antineutrinos \cite{15}, \cite{16}, and \cite{18}. In the Savannah River Plant experiment, the thermal power of the reactor was considered to be $W = 2000$ megawatt, the experimental installation was at a distance of $R = 11.2$ m from the reactor, and the mass of heavy water was $m = 268$ kg. In the Bugey experiment \cite{16}, they were respectively: $W = 2785$ megawatts, $R = 18.5$ m and $m = 267$ kg. For the Rovno experiment \cite{18}, ones consider that $W = 1375 \pm 27$ MW (but in the work \cite{17} ones says that $W = 1195$ MW), $R = 18.1 \pm 1.0$ m and $m = 2985 \pm 3$ kg. 

In the paper \cite{15}, the registration efficiencies averaged over the detector for single-neutron events $\left< \eta \right>$ and for double-neutron events $\left< \eta^{2} \right>$ are considered equal to $0.32 \pm 0.02$ and $0.112 \pm 0.009$, respectively, i.e., the values $\left< \eta^{2} \right>$ and $\left< \eta \right>^{2}$ differ a little enough. The method of finding these efficiencies is not specified. The weighted average number of all events detected per day with one and two neutrons are $R_{1n}=68.3 \pm 16.1$ and $R_{2n}=3.17 \pm 1.83$, respectively.

The main objective of the experiment \cite{15} was to find the value of
\begin{equation}
{\cal R} = \frac{{\rm CCD}_{\rm exp}/{\rm NCD}_{\rm exp}}
{{\rm CCD}_{\rm th}/{\rm NCD}_{\rm th}},
\label{9}
\end{equation}
where CCD (NCD) is the rate of the deuteron disintegration process by a charged (neutral) current of reactor antineutrinos. A few years later, Reines admitted \cite{23} that the intention of the experiment "is based on the fact that the charged current should reflect the presence of an oscillation, if it is in the detectable region." It would seem that the hope on the oscillation detection has became true, because it has been announced \cite{15} that ${\cal R} = 0.38 \pm 0.21$, or $0.40 \pm 0.22$ depending on the theoretical spectrum of reactor antineutrinos. This result was perceived by the scientific community just as an indication of neutrino oscillations (see, for example, \cite{24}, section 29).

It's been almost 20 years. In a number of experiments with reactor antineutrinos, their significant oscillations were not revealed. In agreement with this, the work on the deuteron disintegration at the Bugey reactor \cite{16} has gave the value ${\cal R} = 0.96 \pm 0.23$ for the ratio (\ref{9}). The initial value of the single neutron registration efficiency, found by periodically placing the neutron source $^{252} $ Cf in the center of the detector target, was $\left< \eta \right> = 0.41 \pm 0.01$. The subsequent startup of the Monte Carlo program has gave the value $\left< \eta \right> = 0.29 \pm 0.01$. It was considered that $\left< \eta^{2} \right> = \left< \eta \right>^{2}$. At the same time, in table III of the article \cite{16} there is an undefined "software efficiency"$,$ which is considered equal to $0.471 \pm 0.003$ ($0.444 \pm 0.005$) for the operating (stopped) reactor. The number of events detected per day with one and two neutrons, prior to the application of "software efficiency"$,$ was respectively $R_{1n}=19.3 \pm 0.7$ and $R_{2n}=1.24 \pm 0.16$.

In the experiment at the Rovno reactor \cite{18}, the efficiency of registration of single neutrons in the installation $\eta_{1}$ was calculated by placing sources $^{124}$Sb-Be and $^{238}$Pu-Li (for which the calibration problem is not discussed) in a large number of heavy water points, by finding the registration probability of the neutron, produced in each of these points and by subsequent averaging over them. The result is that $\left< \eta \right> = 0.330 \pm 0.010$ and $\left< \eta^{2} \right> = 0.236 \pm 0.012$. Such a significant difference between the values of $\left< \eta^{2} \right>$ and $\left< \eta \right>^{2}$, $0.236 \pm 0.012$ versus $0.109 \pm 0.007$, could in principle arise under very a large variation of the neutron registration probability under transition from point to point in heavy water, what, apparently, in a reality there is not present.

Further. The number of events registered per day with one and two neutrons in the experiment \cite{18} is $R_{1n}=275 \pm 31$ and $R_{2n}=59 \pm 7$, respectively. Compared to the results at the Savannah Rive and Bugey reactors, the reactor in Rovno has a very high portion of registered events with two neutrons, namely:
$R_{2n}/R_{1n}$ is equal to $0.046 \pm 0.038$ in \cite{15}, $0.064 \pm 0.011$ in \cite{16} and $0.215 \pm 0.050$ in \cite{18}. This situation seems to us to be the most significant summary result of all three experiments discussed.

Note that the ratio $R_{2n}/R_{1n}$ does not depend on the reactor power, on the distance between the plant and the reactor, on the amount of heavy water, and that the declared values of the efficiency of registration of single neutrons in all three experiments are close to each other. Therefore, it is justified to assume that the absence of quenching gas in proportional counters with helium-3 leads to spurious pulses, and that part of the events with one neutron is registered as events with two neutrons. It is quite possible that the portion of spurious double-neutron events depends on all characteristics of proportional counters in their operating state, and that this portion is much higher in the experiment \cite{18} than in the experiments \cite{15} and \cite{16}. The spirious pulses, of course, also lead to an overestimation of the neutron registration efficience, found by means of radioactive sources of neutrons.

There is no doubt that in the absence of a noticeable manifestation of antineutrino oscillations in experiments near a reactor, if they exist at all, the real rate of deuteron disintegration caused by the charged current of the reactor antineutrinos must coincide with the theoretical value $N_{\rm cc}$. Then the equality
\begin{equation}
R_{2n} = \left< \eta^{2} \right> N_{\rm cc}
\label{10}
\end{equation}
can serve as a check of the otherwise found efficiency of registration of events with two neutrons $\left< \eta^{2} \right>$. The experimental rate of deuteron disintegration by neutral currents of reactor antineutrinos $R_ {\rm nc}$ is given by the following equation
\begin{equation}
R_{\rm nc} = \frac{R_{1n}+2 R_{2n}}{\left< \eta \right>} - 2 N_{\rm cc}. 
\label{11}
\end{equation}

Note that all three performed experiments on deuteron disintegration by reactor antineutrinos are valuable above all by lessons which we can take when comparing their results with each other.

\begin{center}
{\large \bf 4. Conclusion}
\end{center}

Setting accurate experiments on the deuteron disintegration by reactor antineutrino is extremely challenging. The preparation and carrying out such experiments should be based on the extensive and diverse methodological and practical base of the SNO, developed and well-proven in setting experiments with solar neutrinos, as well as on new studies of various aspects of neutron registration by proportional counters with helium-3.

I am sincerely grateful to S.P. Baranov for the discussion of problems connected with the present work.

\end{small}
\end{document}